# The fiscal response to revenue shocks

Simon Berset, Martin Huber & Mark Schelker*

University of Fribourg, Switzerland

January, 2021

**Abstract**

We study the impact of fiscal revenue shocks on local fiscal policy. We focus on the very volatile revenues from the immovable property gains tax in the canton of Zurich, Switzerland, and analyze fiscal behavior following large and rare positive and negative revenue shocks. We apply causal machine learning strategies and implement the post-double-selection LASSO estimator to identify the causal effect of revenue shocks on public finances. We show that local policymakers overall predominantly smooth fiscal shocks. However, we also find some patterns consistent with fiscal conservatism, where positive shocks are smoothed, while negative ones are mitigated by spending cuts.

* Department of Economics, University of Fribourg, Boulevard de Pérolles 90, CH-1700 Fribourg, Switzerland. Email: simon.berset@unifr.ch, martin.huber@unifr.ch, mark.schelker@unifr.ch. We thank the Statistisches Amt and the Gemeindeamt of the canton of Zurich for answering our numerous questions efficiently as well as for their invaluable help in our data collection. For comments and discussions we thank Reiner Eichenberger, Sonia Paty, and Christoph Schaltegger as well as participants of the Research Seminar of the University of Basel, the Annual Congress of the European Economic Association (EEA), the International Institute of Public Finance (IIPF), the Swiss Workshop on Local Public Finance and Regional Economics, the workshop of the Swiss Network on Public Economics (SNoPE) at ETH Zurich, the Political Economy and Development Workshop of the University of Zurich, the Ski & Research Workshop of the University of Fribourg, and the Assistentenkonferenz in Innsbruck.

## 1. Introduction

Identifying the drivers of fiscal policy is a daunting task. Observable fiscal outcomes are shaped by many factors such as past policy decisions, the business cycle, financial market conditions, and the institutional, political, and economic environment. All at the same time, these factors might be endogenous themselves to fiscal policy. Even though there is a voluminous academic literature on the drivers of fiscal policy (for recent overviews see, e.g., Alesina and Passalacqua 2016; Yared 2019), it remains notoriously difficult to disentangle the impact of such factors from underlying incentives of decision-makers. The ideal experimental setup would consist of exogenously shifting the budget constraint of a jurisdiction in one fiscal period and observe the induced fiscal response (if any).

Taking a balanced budget as a starting point, unexpected higher revenues (or lower expenditures) result in a surplus, while unexpected lower revenues (or higher expenditures) result in a deficit, *ceteris paribus*. The budget consists of a predetermined part (e.g., entitlement spending, investment spending, interest payments, depreciations, etc.), which cannot be easily adjusted in the short and medium term, and of a discretionary part, which is allocated contemporaneously through a complex political bargaining process among the relevant interests within the institutional setup. Thus, any active policy response in the short run must come from the discretionary part of the budget.

Large fiscal fluctuations—especially unexpected shocks—create opportunities and the potential justifications for decision-makers to use their political leeway to deviate from the *ex ante* budgeted resource allocation in the discretionary part of the budget. We aim to exploit such unexpected and large short-term variations in the tightness of the budget constraint and analyze the fiscal response triggered by it. In order to credibly separate such reactions from larger macro-economic or monetary and fiscal policy dynamics, we focus at the local (instead of the regional or national) level and on revenues from a property transaction tax which is not as closely linked to macro-effects and to the usual tax bases such as income and profit taxes.

We take advantage of arguably (conditionally) exogenous variation of immovable property gains tax (*IPGT*) receipts in the canton of Zurich (Switzerland) and study the expenditure response of local jurisdictions to transitory fluctuations. The *IPGT* is a particularly volatile revenue source. It typically varies around a municipality-specific trend and, from time to time, it is subject to larger shocks. Fluctuations result in short-term (positive or negative) shifts of the budget constraint. The parameters of the *IPGT* are set at the cantonal level, while the proceeds entirely benefit the respective municipality. Municipal decision-makers are aware of the volatility related to *IPGT* receipts. We define *regular flows* as revenue fluctuations that are within a window of what a municipality could expect *ex ante*. In contrast, *positive shocks* and *negative shocks*, are defined as large deviations from this expected trend-window (in our definition, deviations larger than 3 standard errors away from a trend).

For fluctuations to be credibly exogenous for our purposes, they must emanate from idiosyncratic investment and location decisions by private individuals and must be unrelated to municipal public policy or other economic fluctuations such as the business cycle. To identify causal effects, we limit our analysis to large revenue fluctuations, which are typically unexpected by local policymakers and largely driven by individual location and private investment choices, and we purge variation coming from municipality-specific policy as well as economic factors. To do so, we control for underlying trends in the specific tax base and select from a large number of municipality-specific covariates applying causal machine learning methods. We use the post-double-selection method by Belloni, Chernozhukov, and Hansen (2014) based on the LASSO estimator (Tibshirani 1996). Ultimately, identification relies on a conditional independence assumption.

According to traditional, normative public finance theory, the optimal reaction to fiscal fluctuations consists of smoothing them over time. The theory holds that governments should smooth short-term fluctuations and keep tax rates constant to minimize distortions (e.g., Barro 1979; Lucas and Stokey 1987). However, there is a large literature in political economics providing evidence that revenue smoothing is often not the chosen policy. More frequently, political processes feature substantial deficit bias leading to unsustainable public finances in many countries (e.g., Alesina and Passalacqua 2016; Yared 2019; in the local context of the canton of Zurich, Berset and Schelker 2020).

Our baseline results are predominantly in line with normative public finance theory suggesting revenue-smoothing as optimal response to shocks. Only about 20% of positive tax shocks are spent as current expenditures, while no statistically significant effect on current expenditures is observed for negative tax shocks. However, the point estimates of negative shocks are substantially larger than those for positive shocks. These result patterns indicate that the heterogeneity in the expenditure response to negative shocks is large. While there is substantial smoothing (we cannot reject the null hypothesis of perfect smoothing), some part of the sample seems to reacts quite strongly by cutting current expenditures. Hence, some municipalities display a high degree of fiscal conservativeness: they primarily smooth positive shocks, but mitigate negative shocks by spending cuts. Such behavior consists of the direct opposite of deficit bias and is in stark contrast to a politico-economic interpretation.

The rest of the paper is structured as follows. Section 2 reviews the literature and set the relevant theoretical framework. Section 3 formulates testable hypotheses. Section 4 briefly describes the institutional environment of the canton of Zurich. Section 5 presents the functioning of the property gains tax and our approach to distinguish regular fluctuations from shocks. Section 6 discusses the data, the identification strategy and the empirical setup. Section 7 presents results and interpretations. Section 8 concludes.

## 2. Literature review and theoretical framework

The normative theory of tax smoothing initiated by Barro (1979), and further developed by Lucas and Stokey (1987) and Aiyagari et al. (2002) provides a central theoretical prediction of the optimal fiscal response to transitory fluctuations. In order to minimize deadweight losses from taxation, a benevolent social planner would smooth transitory fluctuations in expenditures and revenues through the increase and decrease of debt (or assets). In periods of abnormally positive (negative) fluctuations, the debt-to-income ratio would temporary decrease (increase) but it would remain intertemporally *constant on average*. Permanent structural changes, such as population aging with its implied increases in social security spending, can (optimally) induce adjustments to policy parameters and lead to a new public finance equilibrium.

Several empirical papers analyze the dynamics of fiscal adjustments to revenues and expenditures fluctuations. For instance, Buettner and Wildasin (2006) and Feler and Senses (2017) provide evidence for U.S. municipalities, Buettner (2006) for German municipalities, Solé-Ollé and Sorribas-Navarro (2012) for Spanish municipalities, and Bessho and Ogawa (2015) for Japanese municipalities. Those studies show that municipalities tend to adjust to structural changes and aim to maintain the intertemporal budget balance through adjustments in their current and investment expenditures, as well as through grant transfers.

However, normative theories of optimal fiscal adjustments cannot sufficiently explain the accumulation of public debt in the last few decades. A large politico-economic literature documents and explains systematic tendencies towards deficits ("deficit bias") and the accumulation of public debt (see, e.g., Alesina and Passalacqua 2016; Yared 2019). From this perspective, policymakers are self-interested agents, which optimize according to their private incentives for holding office and fail to internalize the intertemporal consequences of sustained fiscal imbalances. They behave like present-biased agents (e.g., Laibson 1997) with dynamically inconsistent preferences (Yared 2019).

Various political factors and mechanisms contribute to such unsustainable policy making (see, e.g., Alesina and Passalacqua 2016 for a recent overview). Politico-economic theories range from *fiscal illusion*, in which voters do not systematically consider the intertemporal budget constraint (e.g., Buchanan and Wagner 1977); to *political budget cycles*, where voters are imperfectly informed (e.g., Nordhaus 1975; Frey and Ramser 1976; Hibbs 1977; Frey 1978; Rogoff and Sibert 1988; Rogoff 1990); to theories of social conflicts such as the *war of attrition* and *riots*, in which deficit reductions are delayed because the different groups and veto players want to shift the burden of stabilization onto the others (e.g., Alesina and Drazen 1991; Drazen and Easterly 2001; Passarelli and Tabellini 2017); to *public debt seen as a strategic instrument* to constrain future governments in their political decision making (e.g., Persson and Svensson 1989; Alesina and Tabellini 1990; Lizzeri 1999); to *common pool problems* and *legislative bargaining* (e.g., Weingast, Shepsle, and Johnsen 1981; Baron and Ferejohn 1989; Velasco 2000; Battaglini and Coate 2008; Krogstrup and Wyplosz 2010); or to *rent seeking* models, in which policymakers want to extract a maximum of private rents and have to be incentivized by voters to limit rent extraction by keeping them in office (e.g., Acemoglu, Golosov, and Tsyvinski 2008, 2010; Yared 2010; Acemoglu, Golosov, and Tsyvinski 2011).

A smaller part of the politico-economic literature focuses explicitly on *transitory* fluctuations in incomes and government revenues. Several theoretical and empirical contributions emphasize that income and revenue volatility—due to variation in commodity prices (for a review see, e.g., Deaton 1999); in the terms of trade (e.g., Mendoza 1997; Turnovsky and Chattopadhyay 2003; Brueckner and Carneiro 2017); in the tax base (e.g., Gavin and Perotti 1997; Lane 2003); or in foreign aid (e.g., Arellano et al. 2009)—can have an effect on a wide range of outcomes. These outcomes include the impact of income volatility on economic activity (e.g., Fatás and Mihov 2003; Fernández-Villaverde et al. 2015); on armed conflicts (e.g., Dube and Vargas 2013); on corruption, patronage or embezzlement (e.g., Svensson 2000; Caselli and Michaels 2013); or on fiscal policies (e.g., Rodrik 1998; Brueckner and Gradstein 2014). Regarding the latter effects on fiscal policy, two theoretical channels through which fiscal revenue volatility induces inefficient fiscal responses are worth mentioning explicitly.

First, Talvi and Végh (2005) develop an optimal fiscal policy model in which a political distortion causes pressure to increase public spending when governments run surpluses. As budget surpluses become costly, large and anticipated fiscal revenue fluctuations make procyclical fiscal policies optimal for policymakers. As a result, positive fluctuations induce tax reductions and spending increases, while negative ones have the opposite effect. Second, Robinson, Torvik, and Verdier (2017) show that public income volatility makes the implementation of inefficient policies less costly in an environment, in which different groups compete for holding office. The authors use a standard politico-economic model, in which policymakers maximize their re-election probabilities by implementing inefficient policies targeted at their own groups. On the one hand, volatility, and thus, uncertainty in public revenues lowers the benefit of holding office and, therefore, the temptation to implement such inefficient policies decrease. As a result, the reelection probability decreases. On the other hand, as policy inefficiencies are concentrated in the future, inefficient policies become less costly for an incumbent, and the implementation of such policies increase.[1]

In this paper, we focus on the fiscal reaction of local governments to *transitory* and short-term revenue fluctuations from the *IPGT* in the canton of Zurich in Switzerland. We distinguish three types of fluctuations: "regular flows", "positive shocks" and "negative shocks". With this distinction, we investigate whether municipalities react differently to larger fluctuations (shocks) relative to smaller and anticipated "regular" fluctuations, and whether or not positive and negative shocks lead to asymmetric reactions. In contrast to regular flows, the variation in fiscal revenue induced by positive and negative shocks are (1) larger (outside a confidence interval of 3 standard errors around a kernel smoother), (2) unexpected and (3) (conditionally) exogenous from municipal policy decisions.[2]

---

[1] A third mechanism relates to the "voracity effect" proposed by Tornell and Lane (1998, 1999). They develop a theoretical model that emphasizes the critical role of the fiscal process in determining the response to a positive temporary shock. With powerful groups involved in a fiscal process with weak institutions, a "voracity effect" appears and, in equilibrium, the aggregated appropriation is larger than the shock itself. The voracity effect holds that in an environment with weak institutions and powerful groups, an increase in the raw rate of return in the formal economy (e.g., due to a resource windfall) induces incentives of powerful groups to demand more redistribution which increases the tax rate in the formal sector. The increase in the tax burden in the formal sector shifts investments towards the informal sector. This leads to an overall reduction of the growth rate and the over-dissipation of the rents from an increase in the raw rate of return in the formal sector. Strulik (2012) shows that with an elasticity of intertemporal substitution in consumption below unity the voracity effect disappears. Given that the considered shocks in our application do neither result from underlying productivity shocks in the economy nor happen in the context of weak fiscal institution, we do not further explore mechanisms related to the voracity effect.

[2] For empirical evidence of the asymmetric response to positive and negative shocks, see for instance Stine (1994) or Heyndels and Van Driessche (2002). In contrast, Gamkhar and Oates (1996) find evidence of symmetric local reactions to increases and cuts in federal grants.

## 3. Hypotheses: What are the fiscal reactions to shocks?

*Ceteris paribus*, shocks induce variation in the municipal current balance in *t*. Shocks temporarily improve or deteriorate the municipal financial position. The potential fiscal reactions of policymakers to a short-term relaxation or tightening of the budget constraint can fall into four archetypical categories. The first two consist of symmetric responses to positive and negative shocks, while the other two entail asymmetric reactions.

*Smoothing hypothesis:* Following traditional public finance theory, the optimal strategy consists of smoothing unexpected short-term budgetary shocks over time (e.g., Barro 1979; Lucas and Stokey 1987). Positive budget residuals are accumulated and meant to compensate negative ones. Hence, municipalities should neither adjust current expenditures nor current revenues in response to unexpected budgetary shocks. This strategy implies that positive and negative shocks have a symmetric effect. Both types of shocks only affect the current balance in *t*.

*From-hand-to-mouth hypothesis*: Alternatively, jurisdiction could also react sensitively, but symmetrically, to positive and negative shocks, i.e., increase expenditure with a positive shock and decrease it with a negative shock. This, however, requires a strong degree of budgetary flexibility and might induce inefficient fluctuations in the quantity and/or quality of public goods provision. According to Talvi and Végh (2005) and Robinson, Torvik, and Verdier (2017), such budgetary sensitivity is economically less efficient in comparison to smoothing strategies.

*Politico-economic hypothesis:* Positive and negative shocks might trigger asymmetric reactions. While, positive shocks trigger fiscal adjustments, negative shocks do not, and lead to systematic deficits and the accumulation of debt (see, e.g., Alesina and Passalacqua 2016; Yared 2019). Municipal decision-makers might be tempted to use the spending slack in case of a positive shock to allocate these untied resources to specific interest groups, whereas negative shocks do not trigger cuts in spending and/or increases in taxes. Such an asymmetry would indicate that municipalities spend additional resources when available and run a deficit in case of negative shocks.

*Fiscal conservatism hypothesis*: The opposite asymmetry, in which positive shocks are smoothed and negative shocks are mitigated, is also possible. In this case, positive shocks neither affect expenditures nor revenues. They mechanically increase the current balance and capitalize in the stock of assets. However, conservative actors with a deficit aversion might try to avoid deficits at all costs and reduce expenditures or increasing taxes in case of a negative shock. Such an asymmetry would indicate that municipalities mitigate negative shocks, and, over time, accumulate the surpluses from positive shocks in their capital accounts.

## 4. Institutional environment

We focus on variation in the *IPGT* in the canton of Zurich. The tax schedule is defined at the cantonal level, but the tax is levied at and its proceeds are allocated to the municipal level. Municipalities decide upon and provide important public goods autonomously. Moreover, municipalities also decide upon various aspects of their institutional setup.

### 4.1. Municipal fiscal autonomy

The canton of Zurich has 171 municipalities and qualifies as the most fiscally decentralized canton in Switzerland. The ratio of local expenditure relative to the sum of local and cantonal expenditures is about 50%. The municipalities enjoy great autonomy in the definition of public goods and services and the infrastructure they provide. They are responsible for compulsory education at the primary and secondary school levels (30% of current expenses), social assistance (15%), and local health services (5%). Municipalities also provide other public goods and services regarding culture, security, transportation, and the environment. Finally, infrastructure investments account for a significant share of municipal budgets (on average 15% of total annual expenditures). The provision of some of those services is subject to cantonal, sometimes even national standards. However, the municipalities are far from being simple providers of public services defined by upper-layer governments.

On the revenue side, municipalities are subject to the equivalence principle. They primarily finance expenditures with revenues raised through their own taxation of local sources of income and wealth. On average, about half of the municipal revenues come from the direct taxation of natural persons' incomes and firm profits. The overall income and wealth tax scheme is defined by the canton, while municipalities decide on a tax multiplier. An exception is the *IPGT* which is entirely fixed by the cantonal level. Its proceeds amount to an average of about 3.89% of current revenues. The second source of municipal revenues is user charges and fees (18% of current receipts, on average). Unconditional transfers account for only 10% of municipal current revenues, and transfers with a counterpart for 5% on average. This makes the municipalities relatively independent of inter-governmental transfers compared to other local governments worldwide.

### 4.2. Local governance

Municipalities are governed by a "local council", which constitutes the executive and is made up by 5 to 9 members. The local legislative branch is the municipal assembly or, in 13 cases, a local parliament.[3] Local elections are held every four years. Most municipalities only elect the local executive, while citizens constitute the legislative body via municipal assemblies several times a year. Parties play a weaker role at the local level and not all national parties are represented and polarizations is not pronounced.

With the introduction of the new cantonal law on municipalities in 2009, municipalities had to install mandatory budget referendums. A mandatory budget referendum has to be held whenever as a spending proposition lies beyond a certain threshold. The thresholds for recurring expenditures vary between CHF 40'000 and CHF 1 million. Changes to the local income tax multiplier have to be approved by the legislative organ.

### 4.3. Budget formulation and political leeway

Each fall, municipalities prepare a budget for the next fiscal year, coinciding with the calendar year. The planned budget is, on the one hand, a forecast of the financial flows in the forthcoming fiscal year and, on the other hand, the result of the conjunctions of a series of constraints: one part of the budgeted flows is nondiscretionary since it directly results from predetermined expenditure flows, such as entitlements or other spending related to past policy decisions, or cantonal requirements. Another part emanates from local political forces and demands of interest groups.

The revenue-side depends more heavily on forecasts. Municipal authorities formulate their expectations on revenue flows based on their experience, the economic cycle, and other information they might have regarding changes of relevant determinants (e.g., anticipated migration of wealthy taxpayers, announced settlement of a firm, etc.). At the budgetary stage, planned expenditures and forecasted revenues should be close to balance.[4] Hence, budgeted fiscal resources are committed to specific purposes. The planned budget corresponds to an equilibrium outcome that results from politico-economic forces constrained to some extent by the forecasted fiscal revenue. In this equilibrium, not all demands from interest groups can be met and a residual demand remains unsatisfied.

---

[3] Among the 163 municipalities of our studied sample, 9 municipalities have a local parliament over the whole period and one municipality introduced a parliament in 2014.
[4] The municipality law of the canton of Zurich requires that the budgeted current balance must be close to balance. An expenditure surplus can be planned as long as it does not exceed the planned depreciation on the administrative assets plus 3% of the planned tax receipts (Art. 92, Kanton Zürich Regierungsrat 2015).

It is not uncommon for municipalities to see the realized revenue flows deviating—sometimes substantially—from the budgeted values. Positive budget residuals provide additional untied resources, while negative budget residuals result in a lack of resources to finance the budgeted spending. The resulting budget residuals from revenue fluctuations are in the hands of local decision-makers and not *ex ante* determined.[5] Hence, (positive) fiscal shocks may generate budget residuals and provide decision-makers with the leeway to satisfy at least some of this residual demand.

## 5. The immovable property gains tax (*IPGT*) in the canton of Zurich

### 5.1. Setup

In Switzerland, the value-added of immovable properties is subject to taxation.[6] This tax is levied on transactions and not on the annual estimation of the property value. Cantons are in charge of the design of the tax scheme, and the definition of the tax base, the tax schedule, or the distribution of tax receipt between the canton and the municipalities vary across cantons (see Administration fédérale des contributions 2015).

In the canton of Zurich, all real estate transactions, i.e., transactions made by private individuals and by firms, are subject to the *IPGT*. Except for a few exceptions, the private gains made from these transactions are not taxed in other ways. The property gain is calculated as the difference between the purchase price and the selling price, both in nominal terms. The tax scheme in the canton of Zurich is progressive and depends on various parameters. The highest tax bracket is set for gains above CHF 100'000. These are taxed at a 40% rate. To discourage speculation, a surcharge of 50% is added to the tax if the property was held during less than 1 year and 25% if it was held during less than 2 years. A tax deduction is applied for each year the property was held by an owner going from 5 years to 20 years (-3% by year).[7] There is a plethora of additional directives potentially affecting the property gains tax in case of legal persons using the property for core purposes of their commercial activity and for professional property dealers.

---

[5] The annual financial statement is not subject to any balance requirement. Amendments to the accepted planned budget are only required to be mentioned in the annual financial statement.
[6] The definition of « immovable property » is set in the Swiss civil code (Art. 655). Immovable properties include the parcels of land and the buildings thereon, the distinct and permanent rights recorded in the land register, the mines, and the co-ownership shares in immovable property.
[7] The tax code provides the option of postponing the tax payment in few specific cases. Typically, when the transaction concerns a family house and that the household reinvests the product of the transaction in a new home in the same canton, the gain that is reinvested in not subject to taxation. Therefore, the household might ask the postponing of taxation until the second transaction (maximum 2 years).

While the tax scheme is entirely defined by the canton, the fiscal revenue raised from this tax goes exclusively to the municipalities where the transactions took place. On average, the fiscal revenue from the *IPGT* represents 3.89% of the current revenue and 4.49% of the current spending (Table 1, Panel A). Note also that the municipal revenue from this tax is not considered for the resource equalization scheme between municipalities, and thus remains fully at the disposal of the local jurisdiction. The various conditionalities—e.g., the tax rate is a function of the actual profit of a property transaction relative to its initial value at time of purchase (all in nominal terms), the time a property was held by that same owner, and whether or not the selling party purchases property within a certain time window in the canton—make it very difficult to forecast the tax receipt from a specific property transaction, and, at the aggregate level, the total tax receipt of the *IPGT* at the municipal level.

### 5.2. Measuring fluctuations

For each municipality we want to distinguish the anticipated property gains tax receipt from larger and typically unanticipated fluctuations, which we call "shocks". Ideally, we would take the difference between budgeted and realized values. When a realized tax receipt deviates strongly from its *ex ante* forecast, we would consider it as a shock and investigate how the municipality reacted to it. Our strategy follows this intuition with the difference that the budgeted values are not observable and need to be estimated.

For each municipality, we approximate the budgeted tax receipts using a kernel-weighted local linear regression based on an Epanechnikov kernel (Fan 1992; Gutierrez, Linhart, and Pitblado 2003). The method offers several advantages: Local linear regression consists of fitting linear models locally in the neighborhood of specific values of the regressors, with the size of the neighborhood increasing in the bandwidth. It has therefore the simplicity of a standard linear model while being less constraining in terms of linearity assumptions. Here, the bandwidth is specific to each municipality and is calculated according to the rule of thumb (see Silverman 1986), which provides the optimal bandwidth (entailing the minimum mean integrated squared error) for kernel smoothing under normally distributed data. This method has the advantage of permitting for municipality-specific time trends.

The kernel-weighted local polynomial regression is not only technically useful but it is economically meaningful for our purposes. It produces a smoother of the realized tax receipt and an associated standard error for each municipality (Figure 1). For each fiscal period, the smoother can be interpreted as the optimally forecasted *IPGT* receipt per municipality. A municipality that would budget its property gains tax receipt accordingly would, on average, respect intertemporal budget balance.[8]

Figure 1: Distinguishing regular fluctuations from shocks, 4 examples.

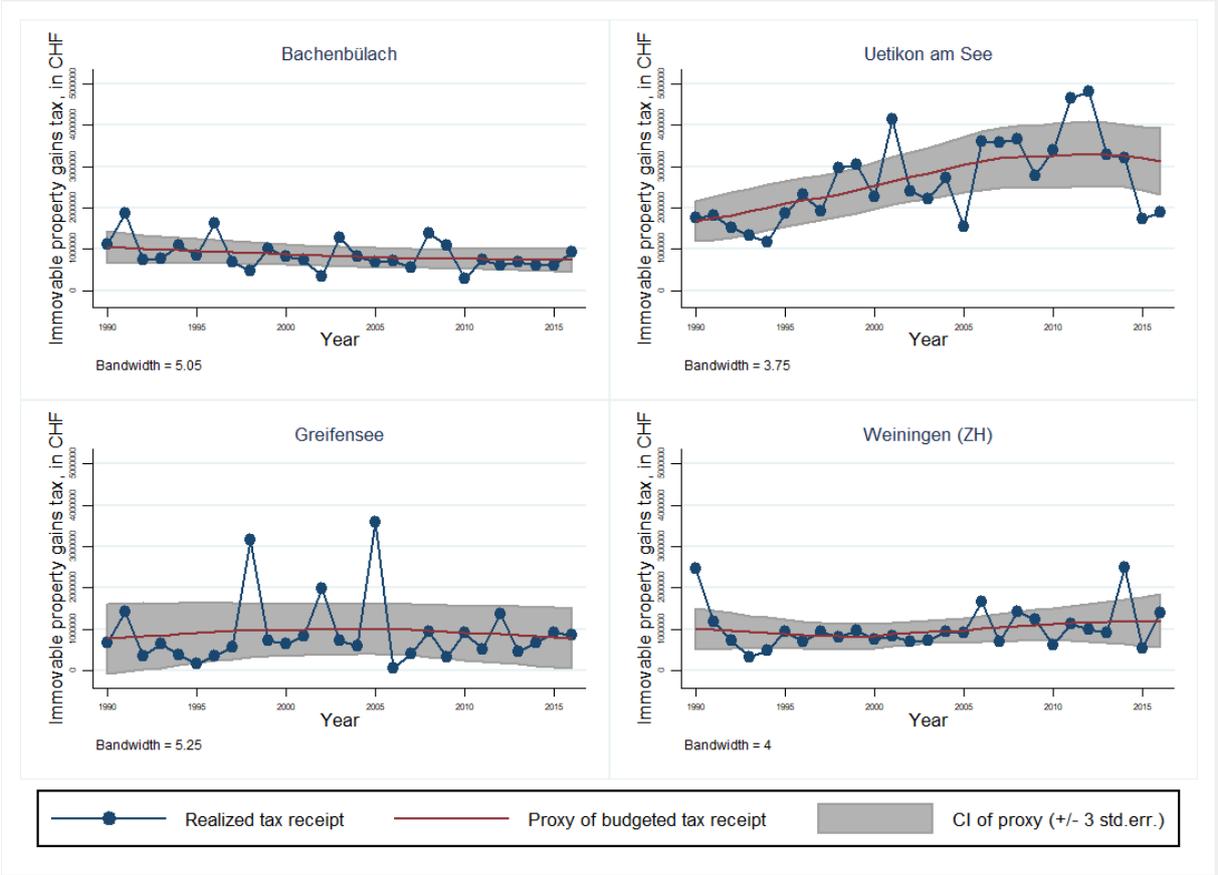

To make sure that our estimation strategy of the budgeted value of the property gains tax approximates actual budgeting, we contacted a sample of 30 municipalities of the canton (of which 15 replied). From the feedbacks received, our estimated smoother seems to be an adequate approximation for the municipality-specific forecasts. Municipalities seem to be well aware of the volatility of this revenue source. The realized revenue flows typically fluctuate around a trend (approximated with the kernel smoother) and only deviate moderately from this *ex ante* forecast. Fluctuations within such a window around the trend will be referred to as *regular flows*. Technically, we define regular flows as the realized tax receipts remaining within a confidence interval around the smoother. Larger fluctuations, outside the confidence interval, constitute the positive and negative shocks.

---

[8] In order to evaluate how precisely the kernel smoother approximates the intertemporal balanced budget path, we compute a measure of relative preciseness for each municipality. We calculate the average of the difference between the realized fiscal receipt and the estimated smoother and express it relative to the average municipal current spending and current revenue. Over the whole sample, the mean of these measures equal -0.05% (Min. -0.97%, Max. 0.99%) and -0.04% (Min. -0.86%, Max. 0.89%) respectively. The smoother is extremely close to the intertemporal balanced budget path. The relative deviation of intertemporal balance does not exceed +/-1% for any municipality.

To define the confidence intervals around the municipal smoothers (and distinguish fiscal shocks from regular flows), we use the standard error associated with the estimated municipality-specific kernel smoothers. By assumption, we consider as "shocks" all realized fiscal receipts that stand below -3 standard errors (negative shocks) or above +3 standard errors (positive shocks). This is an arbitrary definition of a shock. It should strike a balance between defining only larger deviations as shocks while at the same time preserving a sufficient number of observations. We conduct robustness checks with alternative specifications using +/-4 and +/-5 standard errors.

Table 1 presents descriptive statistics of the overall *IPGT* receipts (Panel A), as well as grouped by types of revenue flows (Panels B to D), over the studied period (1990-2016). The *IPGT* receipts amount to an annual average of CHF 1.55 million, which represent 5.8% of the annual average of total municipal current spending and 4.9% of current revenue.[9] With a threshold at +/-3 standard errors, our sample counts 2882 observations classified as regular flows (65.5%), 782 positive shocks (17.8%) and 737 negative shocks (16.7%). Panels B to D provide summary statistics of the difference between the observed *IPGT* receipt and the estimated smoother for each type of revenue flows, i.e. regular flow, positive shocks, and negative shocks. On average, the observed flows that qualify as "regular" (within +/-3 sdt. err. around the smoother) are slightly smaller than the smoother (Panel B). Relative to current spending and revenue, this difference corresponds to about 0.48%. Mechanically, the deltas are much larger for shocks. Positive and negative shocks produce deviations from the smoother that equals about CHF 1 million and CHF 0.6 million on average, respectively.

---

[9] The sample contains 14 negative values mostly concentrated in early 2000. We contacted the cantonal service in charge of municipalities to understand the reason for such negative values. These observations seem to be *ex post* corrections for accounting/attribution errors. There are cases in which the exact amount and timing of an *IPGT* payment had to be corrected to reflect, *ex post*, actual flows.

Table 1: Descriptive statistics of the immovable property gains tax (1990-2016).

|  | Obs. | Mean (Std.err.) | Min. | Max |
|---|---|---|---|---|
| *Panel A: All* | | | | |
| *IPGT*, in 1000 CHF | 4401 | 1,547.54 (2,371.22) | -686.33 | 37,530.87 |
| *IPGT* / current spending, in % | 4401 | 5.80 (5.70) | -8.73 | 114.33 |
| *IPGT* / current revenue, in % | 4401 | 4.93 (4.20) | -9.52 | 62.57 |
| *Panel B: Regular IPGT receipts* | | | | |
| *IPGT* - Smoother, in 1000 CHF | 2882 | -97.96 (464.37) | -5,385.16 | 3,047.00 |
| (*IPGT* - Smoother) / current spending, in % | 2882 | -0.48 (1.67) | -15.08 | 11.13 |
| (*IPGT* - Smoother) / current revenue, in % | 2882 | -0.43 (1.44) | -13.38 | 8.16 |
| *Panel C: Positive shocks (above 3 std. err.)* | | | | |
| *IPGT* - Smoother, in 1000 CHF | 782 | 1,036.71 (1,708.81) | 15.95 | 30,401.20 |
| (*IPGT* - Smoother) / current spending, in % | 782 | 5.33 (5.88) | 0.47 | 80.60 |
| (*IPGT* - Smoother) / current revenue, in % | 782 | 4.23 (3.75) | 0.40 | 38.00 |
| *Panel D: Negative shocks (below -3 std. err.)* | | | | |
| Smoother - *IPGT*, in 1000 CHF | 737 | 612.08 (696.98) | 15.53 | 4,977.35 |
| (Smoother - *IPGT*) / current spending, in % | 737 | 3.11 (2.02) | 0.36 | 22.20 |
| (Smoother - *IPGT*) / current revenue, in % | 737 | 2.79 (1.82) | 0.33 | 17.34 |

Note: Period 1998-2016; without Zürich, Winterthur, and six municipalities involved in local amalgamations.

Importantly, our shock measures are (by definition of the data) asymmetric. While positive shocks have theoretically no upper limit, negative shocks cannot go below zero. In the extreme case, no transactions take place and the resulting fiscal revenue from the *IPGT* is zero. Some small municipalities of our sample collect only very low sums from *IPGT*. Their estimated smoother and its lower bound qualifying negative shocks are, thus, close to zero. This might cause an inflated number of negative shocks. In the robustness section, we exclude municipalities in the lowest decile of the average lower bound (-3 std. err. < CHF 17'800) and the municipalities in the lowest decile of the average smoother relative to current spending (< 1.86%). The results excluding such potential outliers are qualitatively similar. Moreover, the robustness section contains estimates applying a more restrictive definition of what constitutes a shock. We define the bounds of our shock measure to only include fluctuations outside +/- 4 and +/-5 standard errors. Qualitatively the results are similar, but the estimated standard errors of our effects increase due to the smaller number of shocks.

## 6. Data and empirical strategy

### 6.1. Data

Our dataset contains 163 of the 171 municipalities. We exclude the large cities of Zurich and Winterthur, as they represent outliers in various dimensions such as the size, population, demographics, their status of beneficiary of particular transfers for agglomerations, etc. We also exclude 6 municipalities involved in the three amalgamations that took place since 2013. We collect extensive municipal accounting data as well as a wide range of economic, demographic, and socio-economic variables. We were able to collect consistent information for the period from 1990 to 2016. All municipalities use the same accounting model and the same rules apply during the entire period (Direktion der Justiz und des Innern des Kantons Zürich 1984).

As it is standard for public entities in Switzerland, the municipal accounts are organized in three main accounts: the current account, the investment account and the capital account. Our main focus lies on standard public finance measures from the current account, where, by far, most of the discretionary spending originates and most flexibility in the short run exists (e.g., Berset and Schelker 2020). In contrast, investment expenditures are not flexible in the short run, as local investments go through a structured planning and decision-making process. Moreover, investment expenditures are closely linked to local infrastructure and the local property markets. Therefore, measures from the investment accounts play a potentially important role and must be included in the set of covariates. For example, large local infrastructure investments have a very direct effect on property values and, hence, on tax receipts related to property markets. Capital accounts are stock measures and, hence, not suitable in the context of our analysis involving fluctuations in financial flows.

We estimate the impact of shocks on total tax receipts and total current expenditures. The property gains tax enters the municipal "tax receipt" account, which contains all the revenues from all local tax sources (income and property taxes of natural persons, profit taxes of legal persons, etc.). Mechanically, the estimated effect of the property gains tax on this outcome in *t* is expected to be close to 1. If the municipalities do not adjust their tax scheme in response to the property gains tax receipts, the estimated effect should be close to 0 in the subsequent periods. On the one hand, this mechanical effect provides a quality check for our subsequent estimation approach. On the other hand, we can test whether or not municipalities change their tax scheme as a response to a fiscal shock. The current expenditures correspond to the sum of all economically relevant accounts on the expenditure-side of the current account (e.g., personnel spending, operating spending, subsidies, etc.). Accounts that serve pure accounting purposes are excluded from the aggregate (e.g., internal charging).

### 6.2. Understanding the drivers of the immovable property gains tax

For identification, it is important to understand the potential drivers of revenue fluctuations of the *IPGT*. Beside the specificities of the tax scheme, which are set at the cantonal level, numerous factors are likely to affect the annual fiscal revenue from the property gains tax. Mechanically, these are the number of transactions taking place and the value-added in property markets over time. Both of these factors are endogenous to, for example, the location of the municipality (e.g., lake shore, closeness to attractive labor markets), the economic cycle, the growth of real estate markets, migration movements, and other structural changes.

The potentially large number of relevant factors in property markets and the difficulty to forecast them, in concert with the complexity of the tax scheme make also the fiscal revenue itself hard to predict. This is curse and blessing at the same time: From an econometric point of view, a disadvantage is that specifying an unbiased and sparse model is challenging. However, it also makes it difficult for local policymakers to predict revenue fluctuations, which limits anticipation effects and the potential of reverse causality. According to our explorative survey, municipalities forecast the revenue from the *IPGT* primarily on their experience of the previous years (the number of transactions and the average tax receipt from the last 3 to 5 years) and a few other known parameters such as the evolution of land prices in the municipality or the municipal reserve of building area (which are observable to the econometrician). This information allows municipalities to forecast a trend window (which we capture with "regular flows"), but they are unlikely to estimate revenues more precisely.[10]

---

[10] It could be feared that exceptionally large positives fluctuations, due to only a small number of particularly large transactions, might be easy to forecast because such cases might imply long negotiations or prior announcements. However, respondents of our exploratory survey emphasized that even with large and publicly debated

### 6.3. Identification strategy

We are interested in the causal effect of positive and negative fiscal revenue shocks from the *IPGT* on local public finances. Identifying causal effects is not trivial, as policymakers might not only anticipate revenue trends ("regular flows"), but potentially also revenue shocks, and because the shocks might have been triggered by some political factors. In what follows, we address central identification challenges.

*a) Timing*

To capture potential anticipation effects, we introduce the shock measures for two pre-treatment periods. Similarly, a fiscal revenue shock might trigger delayed and/or persistent adjustments over the following budgetary periods. To capture such delayed responses, we include 5 post-treatment periods in our baseline regression (from $t – 2$ to $t + 5$), and up to 10 post-treatment periods in the robustness section (form $t – 2$ to $t + 10$). We are careful to include a sufficient number of post-treatment periods to make sure that any potential effects of a shock have time to fade out. If that were not the case, effectively treated periods (due to a potentially lasting fiscal change given some shock) would end up among the control periods and, hence, bias the estimates. In simple event study setups with a permanent change in the treatment status, this problem is solved by endpoint binning (e.g., Schmidheiny and Siegloch 2020) which, however, is not possible in our setup with one-off and potentially repeated shocks.

*b) Endogeneity to local public policy decisions*

Even though the parameters of the *IPGT* are defined at the cantonal level and local property transactions and their specific timing are largely driven by private decisions (e.g., migration due to a job offer, family reasons, etc.), the potential endogeneity of the *IPGT* fluctuations to local public policy decisions is a direct threat to valid inference. However, changes to local public policy are typically predictable, rarely come as shocks and tend to be persistent at least for some time. Thus, they would change the local trend in property markets and should be picked-up by the smoother. Moreover, most of this variance is likely to be captured by "regular flows", yet, we cannot exclude that some policy changes might also cause larger deviations qualifying as shocks. Ideally, we would like to purge all variation related to local policies. We dispose of a large number of covariates that reflect local institutions such as whether a

---

transactions, it remains usually unclear what specific tax rate (depending on various parameters) would finally apply and when the proceeds actually enter the accounts. The number of transactions, as well as the generated value-added, varies often substantially from their forecasts. Deviations might be positive with more transactions of greater value, or negative, with fewer and smaller transactions than expected.

municipality holds town meetings or has a parliament, whether or not there is a mandatory fiscal referendum, the electoral cycle, or covariates relating to policy outcomes such as infrastructure investments, public goods and services, vertical transfers, and other economic, socio-demographic and other variables.

*c) Endogeneity to the business cycle*

As we are interested in the fiscal response of local policymakers independent of the economic environment, a second source of potentially confounding variation is the local macro-economic cycle. We are able to control for business cycle dynamics captured by the unemployment rate and the tax capacity (essentially the normalized tax base) of natural and legal persons incomes and profits. We can also control for trends in the local property markets. A relatively direct measure of such trends is the trend receipts of the *IPGT*, which we approximate with our kernel smoother. Municipal and time fixed effects and municipality-specific linear and quadratic time trends control for invariable (e.g., lake shore) and slow-moving location effects as well as year specific effects (e.g., financial markets), all of which can affect property markets and migration decisions.[11] Note that national fiscal policy decisions as well as monetary policy affect municipalities similarly and are thus accounted for by time effects.

*d) Covariate selection and the risk of overfitting*

Given the large number of potential covariates as well as the demanding lag-structure, the risk of overfitting is high. Therefore, we opt for the post-double-selection estimator by Belloni, Chernozhukov, and Hansen (2014), which consists of a data-driven process of covariate selection based on LASSO estimation (Tibshirani 1996). The LASSO is a variable shrinkage machine learning method which selects the relevant controls among a large number of covariates. The method fits perfectly our setup with limited degrees of freedom and a risk of overfitting.[12] Ultimately, however, our identification relies on a conditional independence assumption (e.g., Wooldridge 2002).

Following the post-double-selection methodology, we first select the set of control variables that best predict the respective outcome variable. Secondly, we select the variables which best explain our causal variables (this step is repeated for each variable of interest). Third, we estimate the full model using the union of the selected covariates from the two previous steps in a simple OLS regression.

---

[11] Note, however, that it would be problematic to directly control for local property prices, as they are also affected by idiosyncratic private decisions. Controlling for this variation would purge relevant variation in the *IPGT* and, hence, bias the results.
[12] We also implemented estimators based on causal random forests. Given the limited sample size, the method provided only very noisy estimates and proved to be inadequate in our case.

*e) Bad controls*

We need to pay particular attention to the set of covariates from which the algorithm may select. Our goal is to estimate the impact of large variations (shocks) in *IPGT* receipts on particular public finance outcomes. These shocks should be (conditionally) orthogonal to municipal policy decisions and must emanate from independent private decisions that generate unexpectedly high (or low) tax receipts. Hence, controlling directly for parameters reflecting private decisions would absorb part of the relevant variation and would qualify as bad controls (Angrist and Pischke 2009). Therefore, we do not include covariates such as the number of real estate transactions, real estate prices, or migration movements in the pool of potential covariates.

The set of covariates includes, for instance, information on the available building and construction area, investments in transportation and other public infrastructure, and a wide range of economic, political, demographic, and socio-economic municipal characteristics. We also dispose of the necessary information to separate the effect of the shocks from the potential impact of transfers such as equalizations transfers, cantonal grants, and the like. All variables enter the pool of potential covariates with the same temporal structure as our main explanatory variables and they range from $t - 2$ to $t + 5$. Finally, we include municipal linear and quadratic time trends. We chose to never penalize municipal and year fixed effects. All estimations include robust standard errors clustered at the municipal level.

### 6.4. Empirical specification

Based on the kernel-weighted local polynomial regression, we are able to distinguish regular flows from positive and negative shocks in the *IPGT* receipts (*IPGT*). To this end, we generate two binary variables: $Shock_{i,t}^{Pos}$ and $Shock_{i,t}^{Neg}$. $Shock_{i,t}^{Pos}$ equals 1 when the tax revenue is considered a positive shock and $Shock_{i,t}^{Neg}$ equals 1 when the shock is negative (and zero otherwise). From this setup we are able to infer the effect of regular variation within a trend window (+/-3 standard errors of the kernel-weighted local polynomial smoother) as well as the effects of positive and negative shocks (defined as deviations from the kernel smoother), and evaluate the effects against the null-hypothesis of not being different from zero. In the case of shocks, we also estimate whether or not the shocks are statistically different from one another to evaluate the symmetry of the fiscal reaction (see our hypotheses).

We estimate (variants of) the following distributed lag model and use the (conditionally exogenous) *IPGT* shocks to obtain the effect on tax receipts and current expenditures. Our analyses show that potential effects fade out over 5 post-treatment periods.[13]

$$Y_{it}^j = \alpha + \sum_{\tau=-5}^{\tau=2} \beta_\tau IPGT_{i,t+\tau} + \sum_{\tau=-5}^{\tau=2} \gamma_\tau Shock_{i,t+\tau}^{Pos} + \sum_{\tau=-5}^{\tau=2} \delta_\tau Shock_{i,t+\tau}^{Neg}$$
$$+ \sum_{\tau=-5}^{\tau=2} \sigma_\tau IPGT_{i,t+\tau} \times Shock_{i,t+\tau}^{Pos} + \sum_{\tau=-5}^{\tau=2} \rho_\tau IPGT_{i,t+\tau} \times Shock_{i,t+\tau}^{Neg}$$
$$+ \sum_{\tau=-5}^{\tau=2} \varphi_\tau Smoother_{i,t+\tau} + X_{i,t+\tau}\boldsymbol{\theta} + \vartheta_i + \mu_t + \epsilon_{it}, \qquad \text{Eq. 1}$$

with the indices *i* and *t* referring, respectively, to municipalities and years, and $\tau$ reflecting the lag and lead structure for the variables of interest. With the lead and lag structure, we estimate potential anticipation and persistence effects. The index *j* refers to the outcome variables *tax receipts* and *current expenditures* ($Y_{it}^j$). Besides the *IPGT* and shock measures, the specification includes the $Smoother_{i,t}$ and a matrix of LASSO-selected covariates ($X_{it+\tau}$), and municipal ($\vartheta_i$) and time ($\mu_t$) fixed effects.

We always include the $Smoother_{i,t}$, which reflects the municipality-specific trend in *IPGT* receipts according to the kernel-weighted local polynomial regression. It allows us to control for the expected trend revenue flow respecting the intertemporal budget constraint. It is an (*ex post*) optimal intertemporal prediction of the revenue flow and incorporates trends in real estate markets, the business cycle, and other covariates that affect *IPGT* receipts.

Our main variables of interest are the interaction terms: The interaction $IPGT_{i,t} \times Shock_{i,t}^{Pos}$ estimates the marginal effect (in CHF) of a positive shock relative to a regular revenue flow on public finance outcomes in a specific year. The second interaction $IPGT_{i,t} \times Shock_{i,t}^{Neg}$ estimates the marginal effect (in CHF) of a negative shock relative to a regular flow in a specific year. Hence, our estimates are based on the intensive margin of the *IPGT* variation and the effects have a direct monetary interpretation.[14] It is important to note that effects which are specific to a shock year in some municipality (but not specific to the *IPGT* flow) are controlled for by the $Shock_{i,t}$ dummies (positive and negative).

---

[13] In the robustness section we also present results from a specification including 10 post-treatment periods (10 lags) showing that any effects fade out up to the lag 5.
[14] For example, a correlation of 0.75 corresponds to a CHF 0.75 reaction to a CHF 1 of fiscal variation.

## 7. Results and interpretations

For illustrative purposes we first present regression results simply on the total *IPGT* variation, hence, without a distinction between regular flows and shocks. We then proceed to our main analysis and present the results differentiating between regular flows and positive and negative shocks.

We report the results in graphs, where we plot the estimated coefficients and the respective confidence intervals when testing against zero. Each plot relates to one outcome and shows the total estimated impact of regular flows, positive shocks, and negative shocks. The total impact of a shock is the sum of the baseline effect (regular flow) plus the interaction effect, i.e., the $Total\ impact\ of\ positive\ shocks_\tau = \beta_\tau + \sigma_\tau$ from Eq. 1. The 95% confidence intervals around the point estimates test against the null hypothesis of a coefficient not being significantly different from zero.

Below the graph, we document the p-values of further test statistics. First, we present significance tests of the shock coefficients (e.g., positive shock) against regular flows as well as against the other shock coefficient (e.g., negative shock). Secondly, we also present p-values of joint significance F-tests including four post-treatment periods (lag 1 to lag 4). Again, we present such tests for both shocks against regular flows as well as between the two shocks. These tests allow as to distinguish between our four hypotheses: "smoothing", "from-hand-to-mouth", "politico-economic", and "fiscal conservatism".

Following the manifesto by the American Statistical Association (Wasserstein and Lazar 2016; Wasserstein, Schirm, and Lazar 2019) and the group of more than 800 scientists in a Nature comment (Amrhein, Greenland, and McShane 2019), we pay particular attention to the economic relevance of our results, rather than focusing exclusively on statistical significance (McCloskey and Ziliak 1996).

### 7.1. Fiscal response to IPGT receipts

As a starting point, we present regressions on the effect of the total *IPGT* flow on two main public finance variables: total tax receipts and current expenditures. Only in the next step will we distinguish positive and negative shocks from regular flows.

## a. Tax receipts: mechanical effect

We start with regression results on total tax receipts (Figure 2). The revenues from the *IPGT* are contained in this aggregate account including the receipts of all other tax bases such as the income tax from natural and legal persons. This regression informs us directly on two important issues: First, econometrically, it serves as a specification test. Mechanically, the entry of the *IPGT* should show up as a one-to-one relationship, i.e., the coefficient should be one in treatment period *t* and zero in pre- and post-treatment periods. This exercise provides evidence of whether or not other tax bases (such as income and profit taxes) are correlated with the *IPGT*. If there is a correlation different from zero in non-treatment years and/or a correlation different from one in treatment years, the results might point to endogeneity issues. Secondly, if it were the case that the one-to-one relationship did not hold, this regression would quantify the size of the reaction in other tax bases.

Figure 2: Effect of *IPGT* on municipal tax receipts.

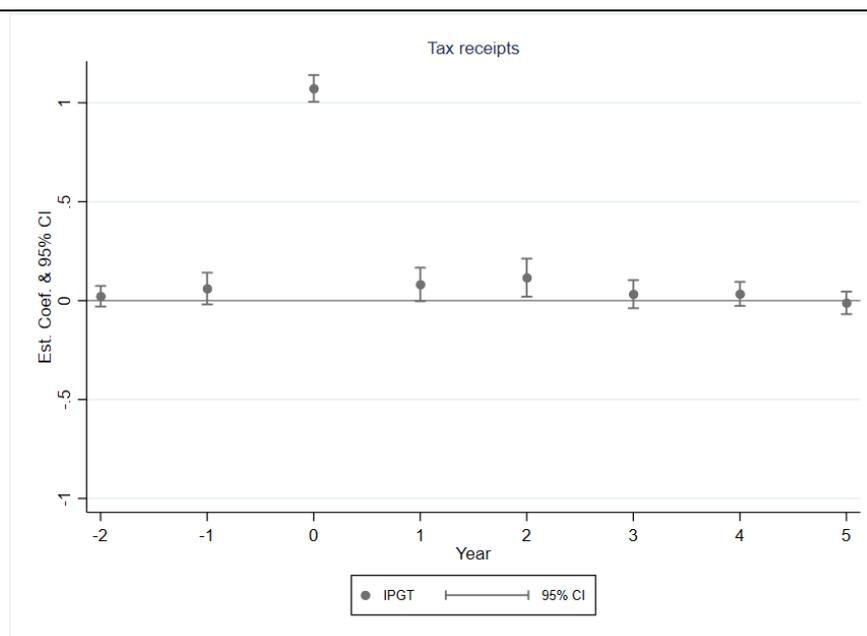

Notes: This figure reports the estimated coefficients of the impact of the immovable property gains tax (*IPGT*) on total tax receipts from the estimation of a distributed lag model according to a variant of equation 1. The 95% confidence intervals around the point estimates test against the null hypothesis of coefficients not being significantly different from zero.

The estimated coefficients displayed in Figure 2 follow quite closely but not exactly a one-to-one relationship between *IPGT* flows and total tax receipts. A CHF 1 inflow of *IPGT* increases total tax receipts by roughly CHF 1.07 in year *t*. This coefficient is significantly different from one at the 5% level (not reported in Figure 2). The two post-treatment effects in $t+1$ and $t+2$ are with 0.08 and 0.11, respectively, also slightly off the zero benchmark. The coefficients are significantly different from zero at the 10% and at the 5% level, respectively. The other coefficients are very close to and not significantly different from zero. Overall, though, the differences to the one-to-one benchmark are rather small. We will further explore the issue in the subsequent analyses distinguishing regular flows from shocks.

b. *Current expenditures*

Figure 3 provides evidence on the current expenditure response to variation in *IPGT* receipts. There is a statistically significant effect in the treatment period $t$ and a significant effect (10% level) in the first post-treatment period $t + 1$ when testing against the null hypothesis of perfect smoothing. Both effects are with 0.19 and 0.18 of similar magnitude. There are neither significant pre-treatment effects or trends, nor any further effects beyond $t + 1$. Hence, a total response of CHF 0.37 per CHF 1 of *IPGT* receipt is found in current expenditures. Considering the small but significant deviations from the one-to-one benchmark for total tax receipts (0.07 + 0.08 + 0.11), the overall adjusted *IPGT* effect on current expenditures amounts to roughly CHF 0.11 per CHF 1 of *IPGT* receipt, i.e., about 11% of *IPGT* receipts are not smoothed.

Figure 3: Effect of *IPGT* on municipal current expenditures.

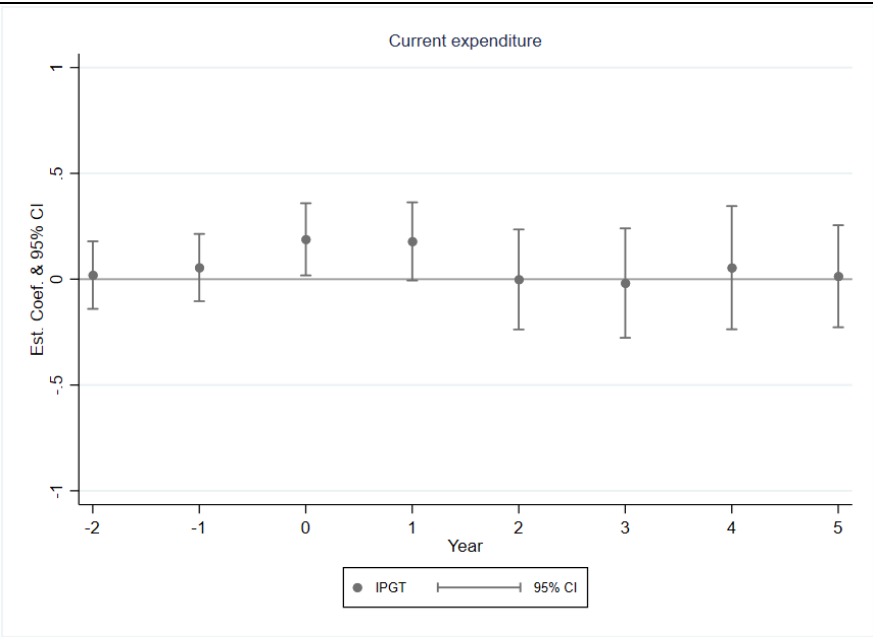

Notes: This figure reports the coefficients of the impact of the immovable property gains tax (*IPGT*) on current expenditures from the estimation of a distributed lag model according to a variant of equation 1. The 95% confidence intervals around the point estimates test against the null hypothesis of coefficients not being significantly different from zero.

## 7.2. Fiscal response to positive and negative shocks: Signs of fiscal conservatism

In our main analysis, we distinguish expected "regular flows" within a window +/-3 standard errors around the kernel smoother from "positive shocks" and "negative shocks", situating outside this window. Our theoretical arguments focused primarily on unexpected shocks. Such shocks shift the budget constraint in the short run and provide political slack in the reaction to the unexpected tax variation. We argued that such shocks are difficult to predict, and are hence more credibly conditionally exogenous than "regular flows". In what follows we provide regression results on total tax receipts and current expenditures which distinguish regular flows from actual shocks.

*a. Tax receipts: mechanical effect*

Once more, this regression informs us directly on the quality of the estimation setup. The *IPGT* enters the municipal accounts through the account "tax receipt" in year *t*, which, *ceteris paribus*, has to be reflected mechanically as a one-to-one relationship in this specific account. Conversely, this regression informs us about a potential violation of this *ceteris paribus* assumption: any important deviation from this one-to-one relationship between *IPGT* and total tax receipts might hint to a potential endogeneity problem. From the perspective of a policymaker, the *ceteris paribus* assumption seems credible *a priori*, because 1) actual shocks are not easily anticipated, 2) *IPGT* rates cannot be adjusted locally, and 3) other tax rates are fairly rigid in the short run as changes in tax rates must be approved by voters or the local parliament. However, there might be forces outside a policymakers' influence which can still challenge the assumption. Consequently, deviations from the one-to-one relationship are unlikely to originate from an active political decision, but from some other underlying variation driving simultaneously the *IPGT* and other tax bases.

All coefficients in the treatment period *t* are close to one and significantly different from zero. However, when tested against one (instead of zero), the positive shock coefficient is statistically different from one at the 5% level in the treatment period *t*; negative shocks and regular flows are not statistically different from one. CHF 1 of shock results in roughly CHF 1.1 in municipal total tax receipts in case of a positive shock, and about CHF 1.07 if in case of a negative shock, thought this difference is not statistically significant.[15] The pre-treatment coefficients are close to and statistically not different from zero. Post-treatment effects are also typically not significantly different from zero. Exceptions are the coefficients estimated for positive shocks which amount to 0.1 in $t + 1$ and 0.13 in $t + 2$, both of which are significant at the 10%. Note that the negative shock coefficient in $t + 1$ is similar in size, but far from any conventional level of statistical significance.

---

[15] To prevent misunderstandings: "negative shocks" are still positive revenue flows (just much lower than expected), and thus we observe positive correlations also in the case of "negative" shocks.

Figure 4: Effect of *IPGT* shocks on municipal tax receipts.

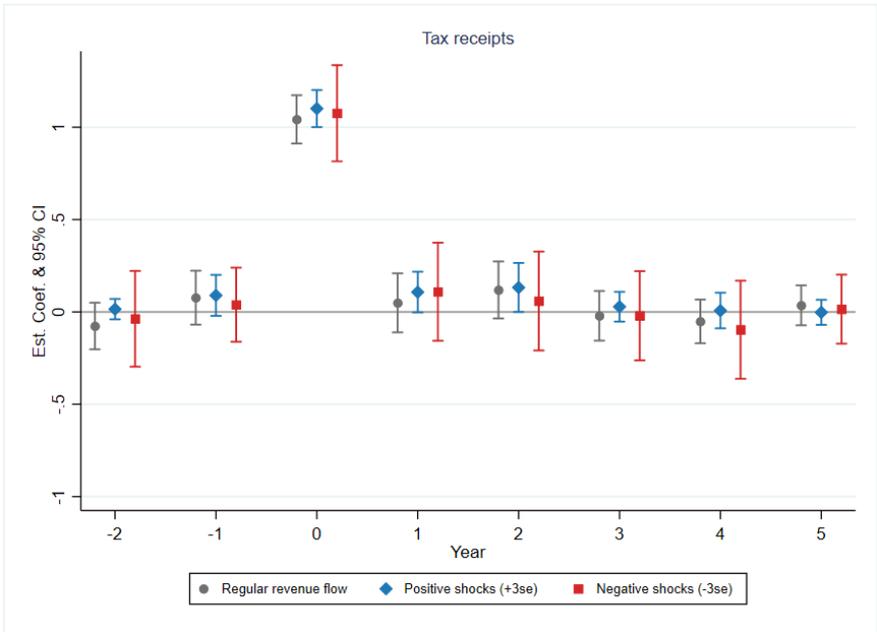

| Timing | t - 2 | t - 1 | t | t + 1 | t + 2 | t + 3 | t + 4 | t + 5 |
|---|---|---|---|---|---|---|---|---|
| **Significance tests: p-values** | | | | | | | | |
| positive vs. regular | 0.05 | 0.73 | 0.07 | 0.33 | 0.75 | 0.33 | 0.12 | 0.26 |
| negative vs. regular | 0.76 | 0.43 | 0.74 | 0.43 | 0.46 | 1.00 | 0.67 | 0.74 |
| positive vs. negative | 0.69 | 0.45 | 0.83 | 0.99 | 0.45 | 0.64 | 0.31 | 0.82 |
| **F-tests of joint significance (L1-L4): p-values** | | | | | | | | |
| positive vs. regular | | | | | 0.37 | | | |
| negative vs. regular | | | | | 0.38 | | | |
| positive vs. negative | | | | | 0.62 | | | |

Notes: This figure reports the coefficients of the impact of regular flows, positive as well as negative shocks of the immovable property gains tax (*IPGT*) on total tax receipts from the estimation of a distributed lag model according to a variant of equation 1. This specification defines regular flows as the variation within +/-3 standard errors around a kernel smoother and the shocks as variation situating outside +/-3 standard errors. The 95% confidence intervals around the point estimates test against the null hypothesis of coefficients not being significantly different from zero (perfect smoothing). The reported significance tests below the graph report p-values of t-tests and joint significance F-tests for lags 1 to 4 when testing either shock against regular flows or against the opposite shock.

In the lower part of Figure 4 we report the p-values of a series of significance tests. In order for us to distinguish among the various hypotheses, we test our shock measures against regular *IPGT* flows and against the respective opposite shock. In the first three rows we report t-tests, in which we test individual coefficients against each other. In the subsequent three rows we document F-tests of joint significance taking together the effect of four post-treatment periods (lag 1 to lag 4).[16] Individually, only very two differences are significant: Positive shocks are significantly different from regular flows in $t-2$ and in $t$, but the differences in coefficients are negligible. None of the joint significance test reach standard levels of statistical significance.

Overall, the expected mechanical effects are present, but positive shocks tend to be correlated beyond the one-to-one relationship. Although, the difference to the one-to-one benchmark in $t$ up to $t+2$ is with about 0.1 rather small and can be taken into account in the subsequent expenditure analyses.

### b. Current expenditures

*Positive shocks*: Both pre-treatment effects are very close to zero and statistically not different from zero (Figure 5). We find statistically significant effects of positive shocks on current expenditures in the treatment period $t$ and the first post-treatment period $t+1$ when tested against zero (perfect smoothing). Both effects indicate an increase in current expenditures of about CHF 0.20 per CHF 1 of positive shock and both are statistically different from zero. None of these effects are statistically different from regular flows and only in $t+3$ positive shocks are statistically different from negative shocks at the 10% level.

*Negative shocks*: None of the negative shock coefficients are significantly different from zero at the 5% level, and only in $t+3$ the coefficient reaches the 10% significance cutoff. Nevertheless, the coefficient sizes are rather large and economically important. In the treatment period $t$ we find neither an economically nor statistically significant effect. This picture changes for the post-treatment period between $t+1$ and $t+3$. Even though none of the coefficients reach the 5% significance level when tested against zero (perfect smoothing), coefficient sizes are with 0.41 and 0.54 rather large. Only in $t+3$, negative shocks are statistically different from regular flows as well as from positive shocks, in both cases at least at the 10% level. None of the joint significance tests indicates statistically significant differences.

---

[16] As shocks fade out by lag 5, we only test for joint significance up to lag 4.

Figure 5: Effect of *IPGT* shocks on municipal current expenditure.

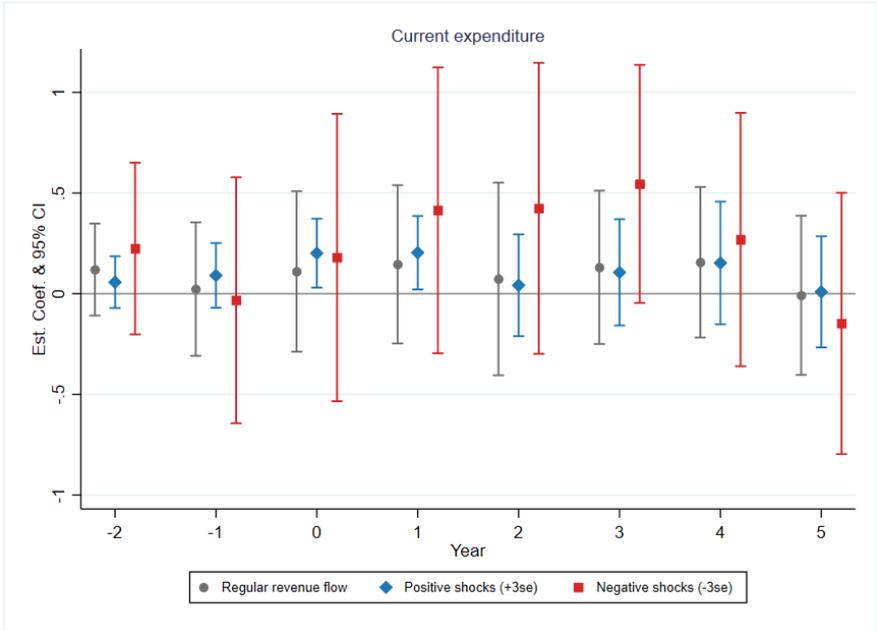

| Timing | t - 2 | t - 1 | t | t + 1 | t + 2 | t + 3 | t + 4 | t + 5 |
|---|---|---|---|---|---|---|---|---|
| **Significance tests: p-values** | | | | | | | | |
| positive vs. regular | 0.44 | 0.54 | 0.52 | 0.68 | 0.83 | 0.85 | 0.97 | 0.85 |
| negative vs. regular | 0.47 | 0.82 | 0.77 | 0.26 | 0.10 | 0.02 | 0.60 | 0.51 |
| positive vs. negative | 0.36 | 0.67 | 0.95 | 0.50 | 0.20 | 0.08 | 0.64 | 0.53 |
| **F-tests of joint significance (L1-L4): p-values** | | | | | | | | |
| positive vs. regular | | | | | 0.91 | | | |
| negative vs. regular | | | | | 0.16 | | | |
| positive vs. negative | | | | | 0.45 | | | |

Notes: This figure reports the coefficients of the impact of regular flows, positive as well as negative shocks of the immovable property gains tax (*IPGT*) on current expenditures from the estimation of a distributed lag model according to a variant of equation 1. This specification defines regular flows as the variation within +/-3 standard errors around a kernel smoother and the shocks as variation situating outside +/-3 standard errors. The 95% confidence intervals around the point estimates test against the null hypothesis of coefficients not being significantly different from zero (perfect smoothing). The reported significance tests below the graph report p-values of t-tests and joint significance F-tests for lags 1 to 4 when testing either shock against regular flows or against the opposite shock.

*c. Result summary*

The findings suggest that positive shocks induce a significant spending response of about CHF 0.40 per CHF 1 (in total) over the treatment period *t* and the first post-treatment period *t* + 1. If we consider that positive shocks are correlated slightly beyond the one-to-one benchmark in the tax receipt regressions, the total *IPGT* effect on current expenditures might only add up to about CHF 0.20. Hence, only very minor portions of *IPGT* receipts are spent on current expenditures in positive shock years and the large majority of *IPGT* receipts is smoothed.

The picture regarding negative shocks is less straight-forward. While the standard errors are generally large and standard levels of statistical significance are almost never reached when testing against zero (perfect smoothing), the post-treatment point estimates in the two years following the treatment ($t + 1$ and $t + 2$) are nevertheless quite substantial and economically relevant. However, due to the rather imprecise estimates, no clear pattern emerges. The lack of statistical significance across all different test statistics indicates that we cannot reject perfect smoothing.

### 7.3. Robustness checks

In what follows, we investigate the robustness of our baseline results by a) excluding outlier municipalities in terms of *IPGT* receipts, b) redefining shocks as variation situating outside of +/-4 and +/-5 standard errors from the kernel smoother, and c) adding additional post-treatment periods (10 lags) to avoid potential estimation bias and to make sure any potentially induced fiscal reaction has sufficient time to fade out.

*a. Excluding outlier municipalities with systematically low property gains tax receipts*

It is important to remember that our shock measures are conceptually asymmetric: While positive shocks have no upper limit, negative shocks have a lower bound at zero (no transactions take place). In this robustness exercise, we exclude 23 municipalities which collect systematically only very minor receipts from the *IPGT* and are situated in the lowest decile.

*Tax receipts*: The mechanical impact on total tax receipts in panel A of Figure 6 is virtually identical to the baseline results in Figure 4.

*Current expenditures*: The results on current expenditures reported in panel B of Figure 6 are qualitatively very similar to the baseline. For positive shocks, there are two marginally significant effects in *t* and *t* + 1 of similar magnitude as in the baseline. Negative shocks have somewhat smaller coefficient sizes and feature even larger standard errors.

Figure 6: Impact of *IPGT* shocks excluding negative outlier municipalities.

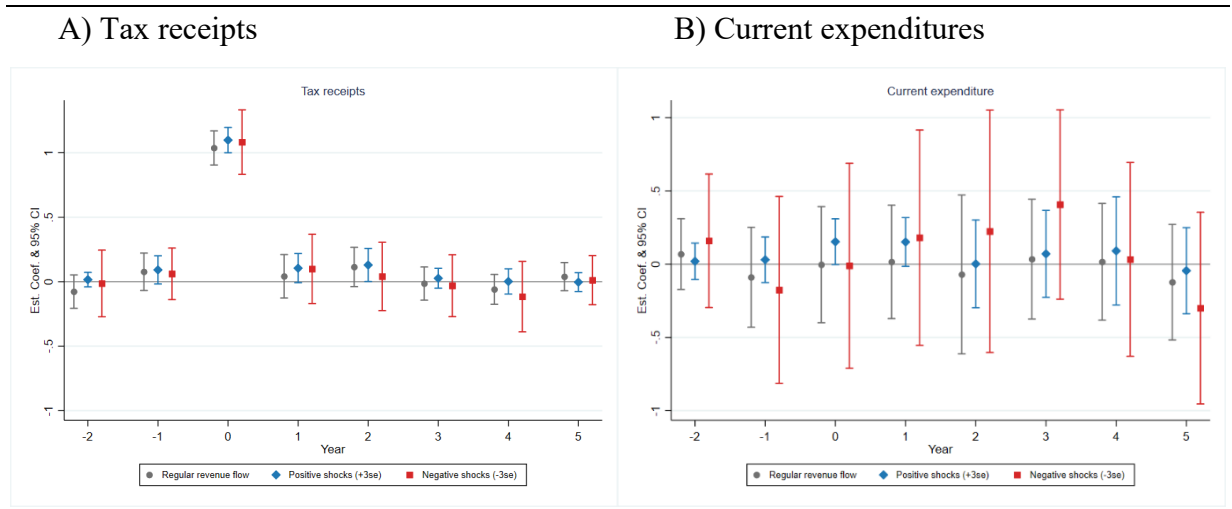

Notes: These figures report the coefficients of the impact of regular flows, positive as well as negative shocks of the immovable property gains tax (*IPGT*) on total tax receipts as well as current expenditures from the estimation of a distributed lag model according to a variant of equation 1. These specifications define regular flows as the variation within +/-3 standard errors around a kernel smoother and the shocks as variation situating outside +/-3 standard errors. The 95% confidence intervals around the point estimates test against the null hypothesis of coefficients not being significantly different from zero. These results exclude 23 outlier municipalities with only very minor tax receipts for the *IPGT* (lowest decile).

b. *Alternative shock thresholds*

The threshold defining a fiscal shock (+/-3 standard errors beyond the smoother) was chosen quite arbitrarily. As robustness checks we conduct the same estimations with shocks defined as deviations of +/-4 and +/-5 standard errors beyond the smoother. These definitions are even more restrictive and fluctuations must be even more extreme to qualify as a shock, which leaves us with much fewer observations.

Figure 7 presents the results using two different definitions of what consists a shock: +/-4 standard errors and +/-5 standard errors beyond the kernel smoother (including again negative outlier municipalities). Overall, the picture remains very similar to the baseline. As can be seen from the estimations on tax receipts (panels A and B), these specifications can replicate the underlying mechanical relationship reasonably well overall, but they do so with lower precision as the number of observations relating to shocks become smaller due to the more severe shock definitions. Especially with the +/-5 standard error definition with observe a level effect in negative shocks. The estimations regarding current expenditures are qualitatively similar to the baseline (panels C and D). The patterns for positive shocks are very similar to the baseline for both shock definitions, although with larger standard errors. The patterns for negative shocks are somewhat more sensitive to the definition of the shock: while the +/-5 standard error definition yields similar but slightly larger point estimates, those with +/-4 standard errors produces somewhat smaller point estimates. However, none of the coefficients reach standard levels of statistical significance.

Figure 7: Impact of *IPGT* shocks under alternative shock definitions.

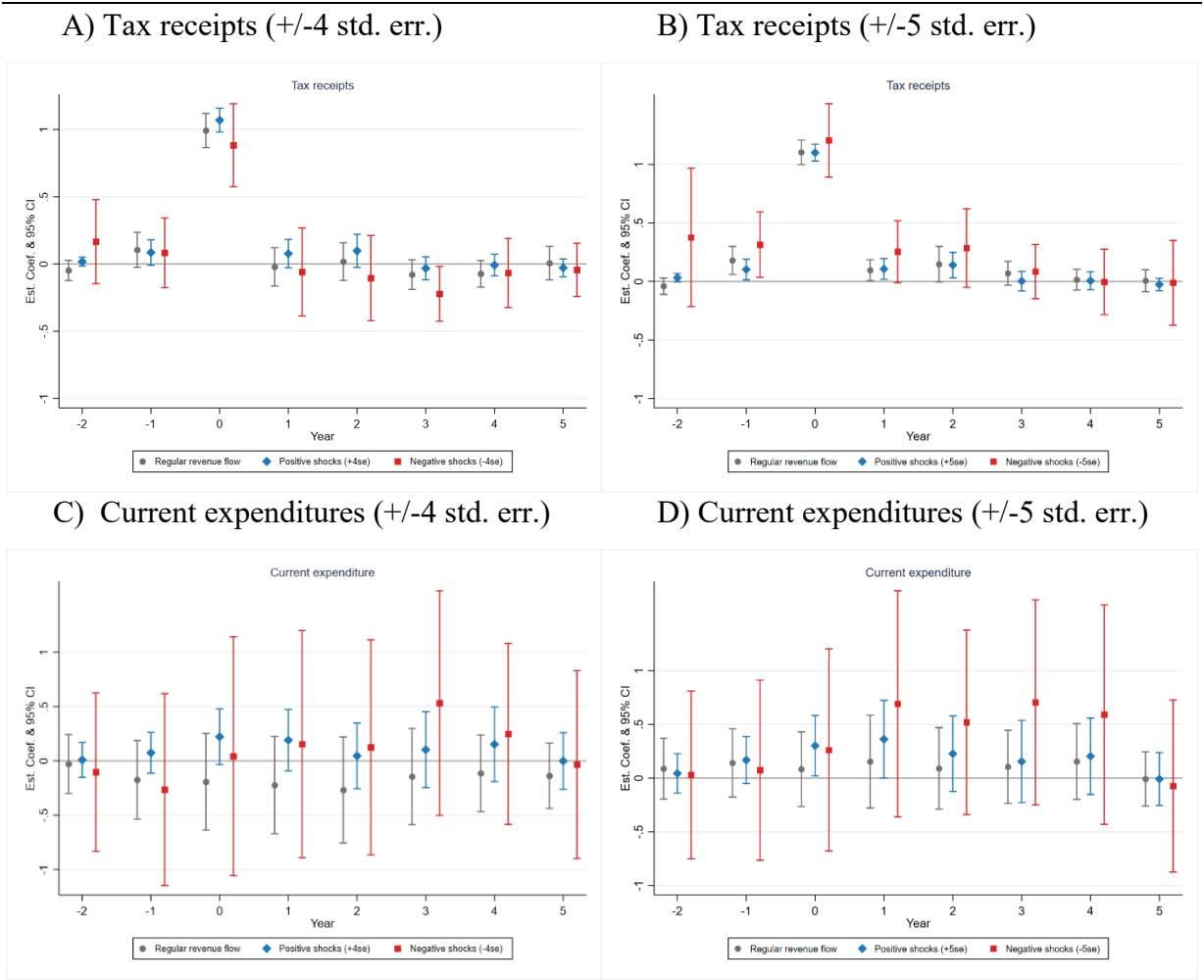

Notes: These figures report the coefficients of the impact of regular flows, positive as well as negative shocks of the immovable property gains tax (*IPGT*) on total tax receipts as well as current expenditures from the estimation of a distributed lag model according to a variant of equation 1. These specifications define regular flows as the variation within +/-4 standard errors around a kernel smoother and the shocks as variation situating outside +/-4 standard errors. The 95% confidence intervals around the point estimates test against the null hypothesis of coefficients not being significantly different from zero.

c. *Additional post-treatment periods (10 lags)*

From an econometric perspective one could be worried that the lag structure in the distributed lag model is misspecified. In a model with finite lags it is not trivial to find the right lag structure. Too few lags bear the risk that some treatment effects in the post-treatment period are not captured by a sufficient number of lags, which could bias the results. Too many lags magnify problems related to multicollinearity and reduce degrees of freedom (as the number of parameters to estimate increase and the number of exploitable periods decrease).

Figure 8 reports specifications including 10 post-treatment periods. Overall, we observe qualitatively very similar patterns as in the baseline with only 5 lags. But more importantly, we observe that by $t + 5$ all estimated effects fade out, be it in the regressions with total tax receipts or current expenditures. The mechanical impact in $t$ equals quite precisely one and none of the other effects are statistically different from zero and there are no important pre-treatment trends or post-treatment effects in any of the *IPGT* variables. The estimated coefficients regarding current expenditure show almost perfect smoothing of positive shocks and somewhat larger point estimates for negative shocks. However, the standard errors are much larger than the baseline results with only 5 lags. For the reasons mentioned above, this is not surprising.

Figure 8: Impact of *IPGT* shocks including 10 post-treatment periods (10 lags).

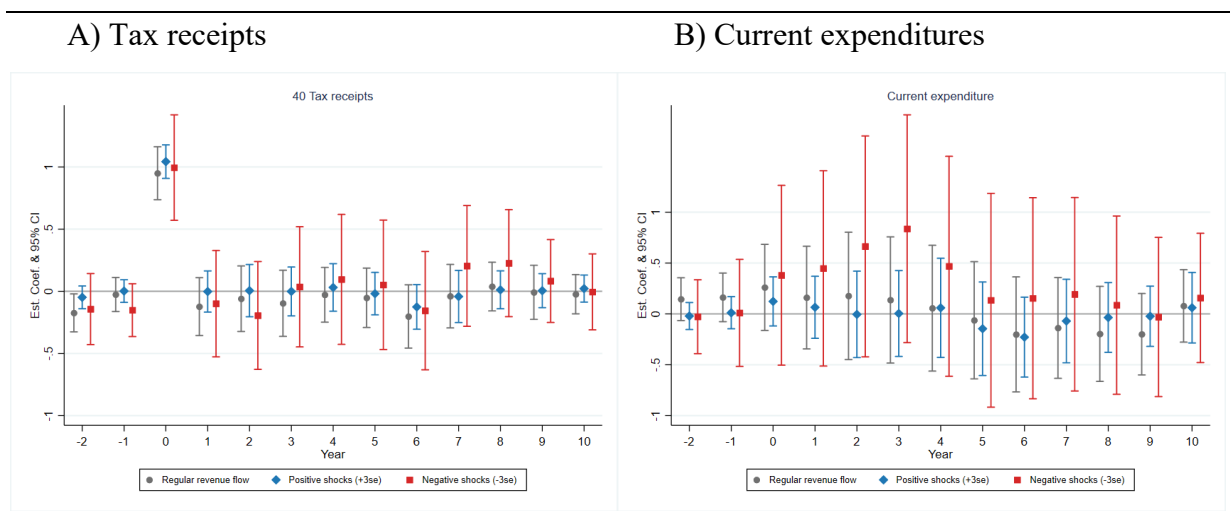

Notes: These figures report the coefficients of the impact of regular flows, positive as well as negative shocks of the immovable property gains tax (*IPGT*) on total tax receipts as well as current expenditures from the estimation of a distributed lag model with 10 lags according to a variant of equation 1. These specifications define regular flows as the variation within +/-3 standard errors around a kernel smoother and the shocks as variation situating outside +/-3 standard errors. The 95% confidence intervals around the point estimates test against the null hypothesis of coefficients not being significantly different from zero.

### 7.4. Interpretation

What is the overall picture of evidence with respect to our formulated hypotheses? Let us first summarize briefly what we found: First, when ignoring effect sizes and only focusing on significance tests, we reject perfect smoothing of positive shocks, while we fail to reject perfect smoothing of negative shocks. At first sight, this might speak in favor of the "politico-economic" hypothesis. In this hypothesis positive shocks are spent, and negative shocks are smoothed, leading to the well-known deficit bias. However, this interpretation is at odds with the overall effect sizes and the absence of significant differences between the shocks. Second, and in contrast, when focusing primarily on effect sizes and ignoring statistical significance, we observe that only about 20% of positive shocks are not smoothed in $t$ and $t + 1$, while at the same time the point estimates of negative shocks are more than twice the size of positive shocks in the post-treatment period (from $t + 1$ to $t + 3$). This would speak in favor of the "fiscal conservativeness" hypothesis, the exact opposite of the previously stated "politico-economic" hypothesis. However, we fail to reject a symmetric reaction between positive and negative shock, which would be required to strongly conclude in favor of either the "politico-economic" or the "fiscal conservativeness" hypothesis.

Therefore, a more nuanced look is necessary: Overall, we have statistically significant evidence that only a small portion of positive shocks are not smoothed and spent in current expenditures. In case of negative shocks, point estimates are large, but confidence intervals always include zero (hence the failure to reject *perfect* smoothing). Nevertheless, confidence intervals asymmetrically include much more territory on the positive side (CI between -0.3 and 1.15 from $t + 1$ to $t + 3$). Hence, i) the failure to reject the null of *perfect* smoothing (when testing against zero) and ii) the large standard errors asymmetrically covering more positive value space, together with iii) the failure to reject a systematic difference in the fiscal reaction to positive and negative shock, suggests that there is considerable heterogeneity in the fiscal response to negative shocks. While there is extensive smoothing in a large part of the sample, there seems to be important mitigation effort to negative shocks (spending cuts) in other parts of the sample.

Taken all together, we conclude that the fiscal reaction to unexpected immovable property gains tax (*IPGT*) shocks in the municipalities of the canton of Zurich is predominantly characterized by *smoothing* and, in some cases, by a sturdy portion of *fiscal conservativeness*.

## 8. Conclusion

Understanding and identifying fiscal behavior of public decision-makers is a daunting task. It requires disentangling underlying political and private incentives from a multitude of endogenous economic and policy factors. In this paper, we take advantage of variation in the immovable property gains tax (*IPGT*), a very volatile source of fiscal revenue in the municipalities of the canton of Zurich. These revenue streams typically vary within a predicted window around a municipality-specific trend, but, from time to time, create budget shocks. These shocks result in short-term shifts (positive or negative) of the municipal budget constraint and provide policymakers the opportunity and justification to use their *ad hoc* political slack to deviate from the budgeted resource allocation in the discretionary part of the budget. Hence, we aim to estimate the effect of fiscal revenue shocks on the spending behavior of local policymakers.

In order to attempt to identify causal effects, we employ causal machine learning techniques. Our double-LASSO variable selection estimates show that, on average, policymakers in the municipalities of the canton of Zurich tend to *smooth* fiscal shocks. However, while only minor parts of positive tax shocks are allocated to increases in current expenditures in $t$ and $t + 1$, point estimates of the impact of negative shocks on current expenditures from $t + 1$ to $t + 3$ are much larger but statistically insignificant. The large heterogeneity in the reaction to negative shocks resulting in large point estimates and large confidence intervals, also suggests that besides fiscal smoothing some part of the sample engages in substantial *fiscally conservative* behavior. In this case positive shocks are primarily smoothed but negative shocks are substantially mitigated via expenditure cuts.

Our results do not point in the direction of a politico-economic hypothesis. This result is in some contrast to the international literature finding widespread evidence for more or less pronounced deficit biases (see, e.g., Alesina and Passalacqua 2016; Yared 2019) and it is in stark contrast to a companion paper (Berset and Schelker 2020) studying a very salient and highly mediatized one-off positive fiscal shock in the same municipalities. Due to the IPO of Glencore on the London Stock Exchange in 2011, municipalities received, on average, a positive fiscal windfall of about CHF 1 million through the cantonal fiscal equalization scheme in 2013. Our causal estimates show that the windfall resulted in large increases in current expenditures (mostly due to expenses for public employees, and subsidies to private individuals) and persistent tax cuts, and caused an increase in municipal debt of about 7.5 times the initial windfall over a period of 5 years before our data end.

One way to reconcile the obvious differences in fiscal behavior could be that different forces are at play: On the one hand, revenue shocks from the appear on a regular basis, affect the own tax base, have to be expected, can be positive as well as negative, and do not cause much media attention. On the other hand, the Glencore shock was truly exceptional, purely positive and clearly non-recurring, and it originated from the tax base of another municipality and only affected a municipality through the fiscal equalization scheme, while creating an enormous amount of media attention. The different characteristics of the shocks (one-off positive *versus* positive and negative as well as potentially recurring) and the difference in salience of the shock might have caused very different reactions and pressures from relevant interest groups.

Svensson, Jakob. 2000. "Foreign aid and rent-seeking." *Journal of International Economics* 51 (2): 437-461.

Talvi, Ernesto, and Carlos A. Végh. 2005. "Tax base variability and procyclical fiscal policy in developing countries." *Journal of Development Economics* 78 (1): 156-190.

Tibshirani, Robert. 1996. "Regression Shrinkage and Selection via the Lasso." *Journal of the Royal Statistical Society. Series B (Methodological)* 58 (1): 267-288.

Tornell, Aaron, and Philip R. Lane. 1998. "Are windfalls a curse?: A non-representative agent model of the current account." *Journal of International Economics* 44 (1): 83-112.

---. 1999. "The Voracity Effect." *American Economic Review* 89 (1): 22-46.

Turnovsky, Stephen J., and Pradip Chattopadhyay. 2003. "Volatility and growth in developing economies: some numerical results and empirical evidence." *Journal of International Economics* 59 (2): 267-295.

Velasco, Andrés. 2000. "Debts and deficits with fragmented fiscal policymaking." *Journal of Public Economics* 76 (1): 105-125.

Wasserstein, Ronald L., and Nicole A. Lazar. 2016. "The ASA Statement on p-Values: Context, Process, and Purpose." *The American Statistician* 70 (2): 129-133.

Wasserstein, Ronald L., Allen L. Schirm, and Nicole A. Lazar. 2019. "Moving to a World Beyond "p < 0.05"." *The American Statistician* 73 (sup1): 1-19.

Weingast, Barry R., Kenneth A. Shepsle, and Christopher Johnsen. 1981. "The Political Economy of Benefits and Costs: A Neoclassical Approach to Distributive Politics." *Journal of Political Economy* 89 (4): 642-664.

Wooldridge, Jeffrey M. 2002. *Econometric analysis of cross section and panel data*. Cambridge, Mass: MIT Press.

Yared, Pierre. 2010. "Politicians, Taxes and Debt." *The Review of Economic Studies* 77 (2): 806-840.

---. 2019. "Rising Government Debt: Causes and Solutions for a Decades-Old Trend." *Journal of Economic Perspectives* 33: 115-140.
38